\documentclass[11pt,dvips]{article}

\usepackage{epsfig,times} 
\usepackage{picinpar}
\makeatletter
\renewcommand{\section}{\@startsection{section}{1}{0in}
	{0.4\baselineskip}{0.1\baselineskip}{\Large\bf}}
\renewcommand{\subsection}{\@startsection{subsection}{2}{0in}
	{0.25\baselineskip}{-\baselineskip}{\large\bf}}
\renewcommand{\subsubsection}{\@startsection{subsubsection}{3}{0in}
	{0.1\baselineskip}{-\baselineskip}{\normalsize\bf}}
\makeatother
\begin{document}

\makeatletter\newcommand{\ps@icrc}{
\renewcommand{\@oddhead}{\slshape{         }\hfil}}
\makeatother\thispagestyle{icrc}
%
%
\hyphenation{hadro-nic brems-strah-lung}

\begin{center}
%
{\LARGE \bf Air Shower Simulations with the AIRES System}
\end{center}

\begin{center}
%
%
{\bf S. J. Sciutto}\\
{\it Departamento de F\'{\i}sica, Universidad Nacional de La Plata,
C. C. 67, 1900 La Plata, Argentina}
\end{center}

\begin{center}
{\large \bf Abstract\\}
\end{center}
\vspace{-0.5ex}
%
%
A report on the characteristics of ultra-high energy air showers
simulated with the AIRES program is presented. The AIRES system
includes a fast simulating program which is an improved version of the
well-known MOCCA program. The AIRES algorithms are briefly described
and a series of results coming from the simulations are analyzed.
%

\vspace{1ex}

%
%
\section{Introduction}
Cosmic rays with energies larger than 100 TeV must be studied --at
present-- using experimental devices located on the surface of the
Earth. This implies that such kind of cosmic rays cannot be detected
directly; it is necessary instead to measure the products of the
atmospheric cascades of particles initiated by the incident astroparticle.

A detailed knowledge of the physics involved is thus necessary to
interpret adequately the measured observables and be able to infer the
properties of the primary particles. This is a complex problem
encompassing many aspects: Interactions of high energy particles,
properties of the atmosphere and the geomagnetic field, etc. Computer
simulation is one of the most convenient tools to quantitatively
analyze such particle showers.

In the case of air showers initiated by ultra-high energy astroparticles
($E\ge 10^{19}$ eV), the primary particles have energies that are several
orders of magnitude larger than the maximum energies attainable in
experimental colliders. This means that the models used to rule the
behavior of such energetic particles must necessarily make extrapolations
from the data available at much lower energies, and there is still no
definitive agreement about what is the most convenient model to accept
among the several available ones (Anchordoqui, Dova, Epele and Sciutto,
1998).

The {\bf AIRES system}%
\footnote{{\bf AIRES} is an acronym for
$\underline{\hbox{\bf AIR}}$-shower
$\underline{\hbox{\bf E}}$xtended
$\underline{\hbox{\bf S}}$imulations.}
(Sciutto, 1999a) is a set of programs to simulate such air showers. One of
the basic objectives considered during the development of the software is
that of designing the program modularly, in order to make it easier to
switch among the different models that are available, without having to
get attached to a particular one.
The well-known {\bf MOCCA}
code created by A. M. Hillas (1997)
has been extensively used as the primary reference when
developing the first version of AIRES (Sciutto, 1997). The physical
algorithms of AIRES 1.2.0 are virtually equivalent to the corresponding
ones from MOCCA (Dova and Sciutto, 1997). The structure of AIRES is
designed to take advantage of present day computers, and therefore the new
program represents an improvement of the MOCCA code, allowing the user to
comfortably perform simulations based on the extensive knowledge on air
shower processes that is contained in MOCCA's source lines.
It is important to remark, however, that the present version of AIRES
(Sciutto, 1999a)
does include modifications to the original algorithms which can alter
the program's output with respect to that from MOCCA. This implies
that both programs are no longer strictly equivalent, even if AIRES's
physical algorithms continue to be largely based on MOCCA's ones.

Another characteristic of ultra-high energy simulations that was
considered when developing AIRES is the large number of particles
involved. For example, a $10^{20}$ eV shower contains about $10^{11}$
secondary particles. From the computational point of view, this fact has
two main consequences that were specially considered at the moment of
designing AIRES: (i) With present day computers, it is virtually
impossible to follow all the generated particles, and therefore a suitable
sampling technique must be used to reduce the number of particles actually
simulated. (ii) The simulation algorithm is CPU intensive, and therefore
it is necessary to make use of a series of special procedures that will
provide an adequate environment to process computationally long tasks.

The particles taken into account by AIRES in the simulations are: Gammas,
electrons, positrons, muons, pions, kaons, eta mesons, nucleons,
anti-nucleons, and nuclei up to $Z=26$. Electron and muon neutrinos are
generated in certain processes (decays) and accounted for their energy,
but not propagated. The primary particle can be any one of the already
mentioned particles, with energy ranging from several GeV up to more
than 1 ZeV ($10^{21}$ eV).

\begin{figwindow}[4,r,%
{\mbox{\epsfig{file=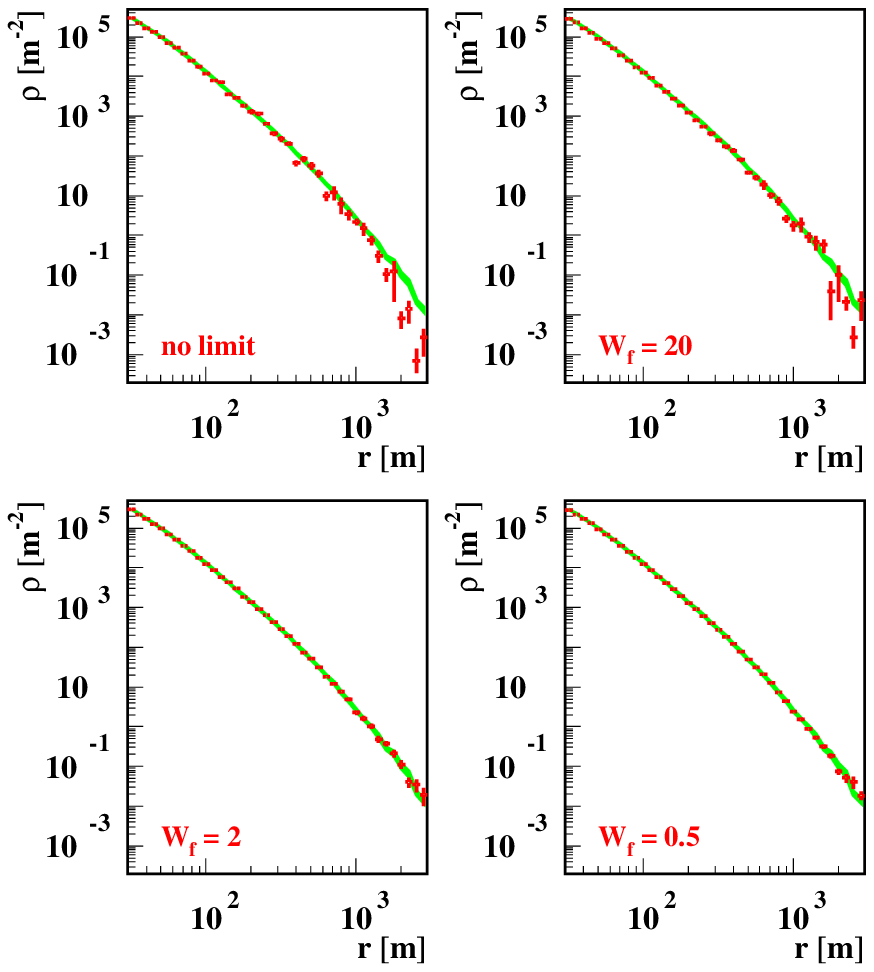}}},%
{Effect of the AIRES extended thinning algorithm on the fluctuations of
the lateral distribution of electrons and positrons.\label{FIG:wf1}}]
Among all the physical processes that may undergo the shower particles,
the most important from the probabilistic point of view are taken into
account in the simulations. Such processes are: (i) {\em Electrodynamical
processes:\/} Pair production and electron-positron annihilation,
bremsstrahlung (electrons and positrons), knock-on electrons ($\delta$
rays), Compton and photoelectric effects, Lan\-dau-Pomeranchuk-Migdal
(LPM) effect and dielectric suppression. (ii) {\em Unstable particle
decays,\/} pions and muons, for instance. (iii) {\em Hadronic
processes:\/} Inelastic collisions hadron-nucleus and photon-nu\-cleus,
sometimes simulated using an external package which implements a given
hadronic interaction model (The current version of AIRES includes
links to SIBYLL (Fletcher {\em et. al.,\/} 1994) and QGSJET (Kalmykov,
Ostapachenco and Pavlov, 1994) hadronic interaction models.). Photonuclear
reactions. Nuclear fragmentation, elastic and inelastic. (iv) {\em
Propagation of particles:\/} Curved Earth geometry, losses of energy in
the medium (ionization), multiple Coulomb scattering and geomagnetic
deflections.

Some of the propagating algorithms and related procedures have recently
been revised, and the impact on the shower observables has been
studied in detail. The results of such analyses have been reported
elsewhere, in particular, in the cases of the LPM effect (Cillis,
Garc\'{\i}a Canal, Fanchiotti and Sciutto, 1998) and the external hadronic
models (Anchordoqui, Dova, Epele and Sciutto, 1998; Anchordoqui, Dova and
Sciutto, 1999).
\end{figwindow}

\section{Ultra-high-energy showers simulated with AIRES.}

Ultra-high energy showers are commonly analyzed in a shower per shower
basis. In the case of simulated data this generally implies generating very
detailed data sets to avoid excessive artificial fluctuations due to the
sampling algorithm, commonly named {\em thinning algorithm.\/}

The AIRES thinning algorithm is an extension of the classical Hillas thinning
algorithm (Hillas, 1981). The AIRES extended thinning algorithm (Sciutto,
1999a) permits limiting the maximum statistical weight a particle entry can
have by means of an external parameter named {\em weight limiting
factor,\/} $W_f$. The limit $W_f\to\infty$ corresponds to the standard
Hillas algorithm while $W_f\to 0$ completely disables rejection of
particles (full sampling).

The plots in figure \ref{FIG:wf1} illustrate how the fluctuations in the
electron and positron density (the data correspond to $10^{19}$ eV
vertical proton showers, ground level at 1400 m.a.s.l) behave when the
weight limiting factor is modified. The first plot --labelled ``no
limit''-- corresponds to simulations performed with the standard Hillas
thinning algorithm ($W_f\to\infty$) and $E_{\mathrm{thin}} =
10^{-5}\,E_{\mathrm{prim}}$ ($10^{-5}$ relative thinning level). The
lateral distribution for $10^{-7}$ relative thinning level (also with
$W_f\to\infty$) is also displayed in the form of a continuous shaded band
($\hbox{\epsfig{file=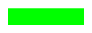}}$) whose width corresponds to the mean
value plus or minus one RMS error of the mean. The number of showers is 25
in each case.

The remaining plots of figure \ref{FIG:wf1} correspond, respectively, to
the cases $W_f=20$, $2$ and $0.5$ ($E_{\mathrm{thin}} =
10^{-5}\,E_{\mathrm{prim}}$). The shaded band corresponding to $10^{-7}$
Hillas algorithm --mentioned in the previous paragraph-- is repeated in
all the plots for comparison.
These graphs illustrate clearly how the fluctuations due to the sampling
method diminish monotonically with $W_f$. In the case $W_f=0.5$ they are
of the same order than those of the $10^{-7}$ case, for all the lateral
distances plotted, that is, for $r\le 2700$ m. Notice also that in this
case, the average CPU time per shower for the simulations performed with
$10^{-5}$ relative thinning level with $W_f=0.5$ is roughly 11 times
smaller than the CPU time necessary to generate one $10^{-7}$ thinned
shower.

\begin{figure}
\[
\begin{array}{cc}
\hbox{\epsfig{file=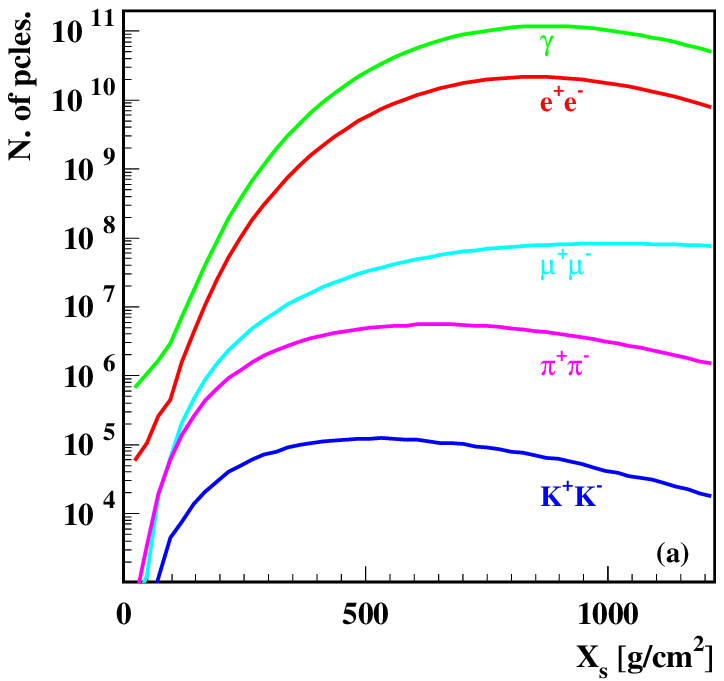}} &
\hbox{\epsfig{file=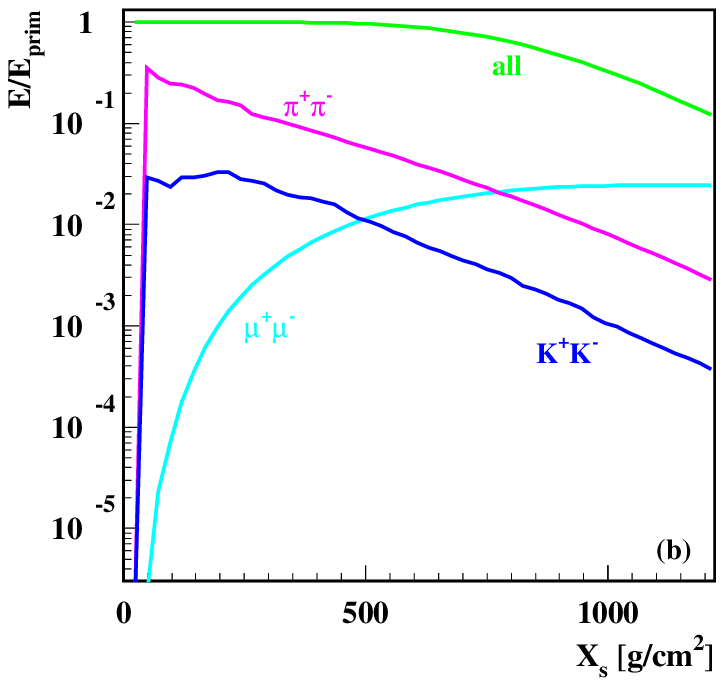}}\\*[-7mm]
\end{array}
\]
\caption{Longitudinal development of a 
{\em single\/} $3\times 10^{19}$ eV proton shower simulated with
AIRES. In (a) the number of different particle kinds is plotted versus the
slant atmospheric depth, while the curves in (b) represent the
corresponding total energies (energy longitudinal
development), in units of the primary energy.\label{FIG:longi8}}
\end{figure}

\begin{figwindow}[0,l,%
{\mbox{\epsfig{file=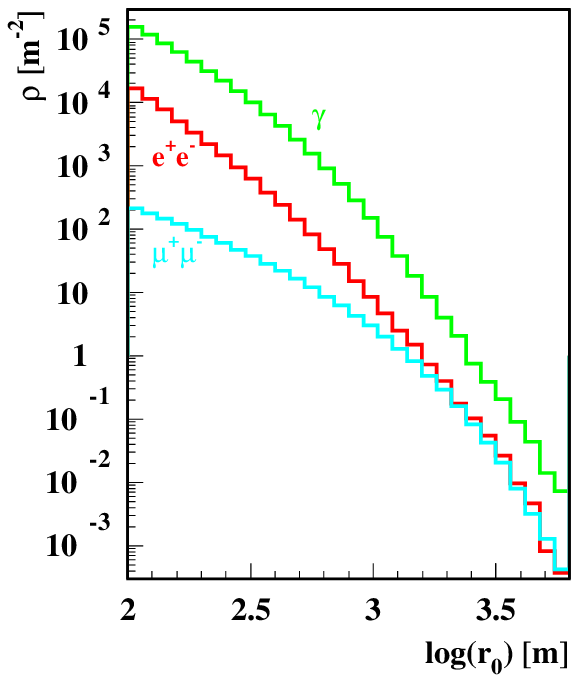}}},%
{Gamma, electron and muon lateral distributions corresponding to a {\em
single\/} $3\times 10^{19}$ eV proton shower simulated with
AIRES.\label{FIG:ld8}}]
When a relatively low thinning energy is combined with a finite weight
limiting factor, the artificial fluctuations due to the statistical
sampling method can be dramatically diminished. Such kind of simulations
may require large amounts of computer time in certain cases, but they can
give a clear enough signal, suitable for detailed simulations of the
response of ground detector arrays.

The lateral distributions of gammas, electrons and positrons, and muons
corresponding to a {\em single\/} $3\times10^{19}$ eV inclined (zenith
angle 45 deg) proton shower are represented in figure \ref{FIG:ld8}.  The
plots correspond to the shower front plane lateral distributions, and
$r_0$ represents the distance to the shower axis measured along this
normal plane. The details about the projection procedure will be presented
elsewhere (Sciutto, 1999b). The lateral distributions show a low level of
noise in all the range $r_0\le 6360$ m, as displayed in figure
\ref{FIG:ld8}. Notice that the maximum value of $r_0=6360$ m, corresponds
to distances to the core up to 9 km, measured along the ground plane.

Global shower observables are also retrieved with a very clear signal.
Figure \ref{FIG:longi8} displays the longitudinal development of this
single shower being considered.
\end{figwindow}

Figure \ref{FIG:longi8}a contains plots of the number of particles versus
the slant atmospheric depth for the cases of gammas, electrons and
positrons, muons, pions and kaons. The shower maximum (all charged
particles) is located at 847 g/cm$^2$; notice that this shower was
simulated using the SIBYLL hadronic package (see Anchordoqui, Dova and
Sciutto, 1999).

Figure \ref{FIG:longi8}b displays the total energy associated to various
particle kinds (divided by the primary energy), plotted as a function of
the slant depth. The curve labelled ``all'' corresponds to the energy of
all particles, and equals 1 at the beginning of the shower development,
indicating that in this part of the shower development the medium energy
losses are very small, and consequently the sum of the energies of the
secondary particles is constant. Then, the continuous energy losses become
more important and the total energy fraction diminishes monotonically
arriving to 0.11 at ground level.
The plots for the hadronic secondaries (pions and kaons) clearly indicate
the position of the first interaction (36.7 g/cm$^2$). Notice that in the
early stages of shower development these particles carry a substantial
fraction of the available energy, and that a part of this energy is passed
to the muons (coming from pion decays), as long as the shower evolves.

\section{Final remarks}

By means of representative examples we have presented some of the main
features of the AIRES system for air shower simulations. For more detailed
information see Sciutto (1999a).

The AIRES system is still evolving; every new release of AIRES represents
a new step towards a complete and more reliable air shower simulation
system. There are still many things pending implementation, for example,
\u Cerenkov photon simulation; multiple primary showers; alternative
atmospheric models; links to other external packages, especially hadronic
models; etc. It is planned to progressively incorporate such additional
features. The current status of the AIRES system can always be checked at
the AIRES Web page: {\ttfamily\bfseries
www.fisica.unlp.edu.ar/auger/aires}.

\section*{Acknowledgments}
The author is indebted to C. Hojvat and all the members of the Auger group
at FERMILAB (Batavia, Illinois, USA) for their kind hospitality.

%
%
%
%
%
%
\vspace{1ex}
\begin{center}
{\Large\bf References}
\end{center}
%
\noindent
Anchordoqui, L., Dova, M. T., Epele, L. N., \& Sciutto, S. J. 1998, preprint
hep-ph/9810384, Phys. Rev. D (1999) (in press).\\
Anchordoqui, L., Dova, M. T., \& Sciutto, S. J. 1999, these proceedings.\\
Cillis, A., Garc\'{\i}a Canal, C. A., Fanchiotti, H.,
\& Sciutto, S. J. 1998, preprint astro-ph/9809334, Phys. Rev. D (1999)
(in press).\\
Dova, M. T., \& Sciutto, S. J. 1997, Pierre Auger technical note
GAP-97-053.\\
Fletcher, R. T., Gaisser, T. K., Lipari, P., \&
 Stanev, T. 1994, Phys. Rev. D, 50, 5710.\\
Hillas, A. M. 1981,  Proc. of the Paris Workshop on
Cascade simulations, J. Linsley and A. M. Hillas (eds.), p 39.\\
Hillas, A. M. 1997, Nucl. Phys. B (Proc. Suppl.) 52B, 29.\\
Kalmykov, N. N., Ostapchenko, S. S., \& Pavlov A. I. 1994,
      Bull. Russ. Acad. Sci. (Physics), 58, 1966.\\
Sciutto, S. J. 1997, Pierre Auger technical note GAP-97-029.\\
Sciutto, S. J. 1999a, AIRES user's guide and reference manual, version
2.0.0, Pierre Auger technical note GAP-99-020.\\
Sciutto, S. J. 1999b, in preparation.

\end{document}